\documentclass[aps,prl,twocolumn,notitlepage,superscriptaddress,nogroupedaddress]{revtex4-1}

\usepackage{xcolor,graphicx}
\usepackage{amssymb,amsmath,graphicx}
\usepackage[utf8]{inputenc}
\usepackage{comment}
\usepackage{braket}
\usepackage[normalem]{ulem}
\usepackage{hyperref,url}

\begin{document}

\title{Third law of thermodynamics and the scaling of quantum computers}

\author{Lorenzo Buffoni}
\affiliation{PQI -- Portuguese Quantum Institute, Portugal}

\author{Stefano Gherardini}
\affiliation{CNR-INO, Area Science Park, Basovizza, I-34149 Trieste, Italy}
\affiliation{LENS, University of Florence, via G. Sansone 1, I-50019 Sesto Fiorentino, Italy}
\affiliation{PQI -- Portuguese Quantum Institute, Portugal}

\author{Emmanuel Zambrini Cruzeiro}
\affiliation{Instituto de Telecomunica\c c\~oes, Lisbon, Portugal}

\author{Yasser Omar}
\affiliation{Instituto Superior Técnico,  Universidade de Lisboa, Lisbon, Portugal}
\affiliation{Centro de Física e Engenharia de Materiais Avançados (CeFEMA), Physics of Information and Quantum Technologies Group, Portugal}
\affiliation{PQI -- Portuguese Quantum Institute, Portugal}

\begin{abstract}
The third law of thermodynamics, also known as the Nernst unattainability principle, puts a fundamental bound on how close a system, whether classical or quantum, can be cooled to a temperature near to absolute zero. On the other hand, a fundamental assumption of quantum computing is to start each computation from a register of qubits initialized in a pure state, i.e., at zero temperature. These conflicting aspects, at the interface between quantum computing and thermodynamics, are often overlooked or, at best, addressed only at a single-qubit level. 
In this work, we argue how the existence of a small, but finite, effective temperature, which makes the initial state a mixed state, poses a real challenge to the fidelity constraints required for the scaling of quantum computers. Our theoretical results, carried out for a generic quantum circuit with $N$-qubit input states, are validated by test runs performed on a real quantum processor. 
\end{abstract}

\maketitle

Large-scale quantum computers represent the ultimate frontier in information processing, with the objective to obtain quantum advantage in solving computational problems that classical computers cannot address in any feasible amount of time \cite{AruteNature2019,WuPRL2021}. So far, one of the biggest obstacles to this endeavour has been noise, which is responsible for the decay of quantum coherence and correlations \cite{KuengPRL2016,NielsenQuantum2021} in quantum states, especially pure states which are notoriously hard to preserve. Quantum error correction protocols \cite{Raussendorf2012PTRSA,Devitt2013RPP}, assisted by the statements of the quantum threshold theorem \cite{Knill1998Science,Aharonov2008SIAM}, can help in overcoming quantum state degradation. However, experiments on existing devices \cite{ibmq,sete2016functional,WrightNatureComm2019} still lack the high-fidelity required for error correction. For this reason, the analysis of thermodynamic and energetic resources has recently emerged in the literature as an useful tool to study the fundamental limits of quantum computation, with several implications on quantum gates \cite{cimini2020npjQI,stevens2021arXiv}, quantum annealers \cite{buffoni2020QST,campisi2021PRE} and quantum error-correction \cite{fellous2020arXiv}.  

In the following, we focus on thermodynamic limits for quantum state preparation, and on their consequences in obtaining high fidelity in multi-qubit quantum registers. The very existence of pure states and the limits to their preparation have to face Nernst’s unattainability principle, also known as the third law of thermodynamics \cite{nernst1906beziehungen}, stating that cooling a physical system to the ground state ideally requires infinite resources. Since pure states can be brought to the ground state (and vice-versa) by means of finite-cost transformations, i.e., unitary operations, the preparation of pure states necessarily involves an infinite resource cost to abide to the third law. This issue has been recently brought to light in the quantum thermodynamics community with implications to quantum measurement \cite{guryanova2020ideal}, purification \cite{ticozzi2014quantum} and cooling \cite{taranto2021landauer}. The simplest and most fundamental case of state preparation is the initialization of a qubits register to the computational state $\ket{00...0}$ by means of the operation denoted as \emph{reset}. Single-qubit reset has been investigated in numerous platforms, some of which are: solid state, such as silicon \cite{Pla2012} or rare-earth ion-doped crystals both in spin ensembles \cite{Longdell2004,Lauritzen2008} and single ions \cite{Utikal2014}, NV centers in diamond \cite{Baier2020}, superconducting qubits \cite{Johnson2012,Riste2012a,Partanen2018,Magnard2018}, microwave photons \cite{Pierre2014,Tan2017}, and trapped ions \cite{Myerson2008,Burrell2010,Schindler2011}. However, $\ket{00...0}$ being a pure state, it is subject to the thermodynamic constraint originated by the Nernst’s principle.

In this work, we show that in real-world quantum computers there exist a thermodynamic limit to the initialization of multi-qubit registers, and consequently to the preparation of pure quantum states, that has practical implications on the scaling of quantum computers. In fact, although the reset (or initialization) of single-qubits has been realized with high-fidelity (even above $99.9\%$) \cite{HartyPRL2014}, we will analytically prove and verify on a real device that even a small initialization error on a multi-qubit register may dramatically reduce the preparation fidelity of a multi-qubit state. We argue that, to scale beyond the actual devices, substantial efforts are needed to improve the quality of initialization of multi-qubit registers. 

\paragraph{Fidelity scaling.--}

The usual assumption in quantum computation is to initialize the qubits register in the computational state $\ket{00...0}$ and then evolve it with an arbitrary unitary operation. Here, we want to investigate how this operation is affected by an imperfect preparation of the initial $\ket{00...0}$ register. Using the formalism of density matrices, the initial $N$-qubit pure state (target state) one wishes to initialize in a quantum computer is 
\begin{equation}
    \sigma_0 \equiv  
    \begin{pmatrix}
    0 & 0 \\
    0 & 1 \\
    \end{pmatrix}^{\otimes N}
\end{equation}
that, by definition, is a zero temperature state. However, from the Nernst’s unattainability principle we are bound to prepare states that have an arbitrary small, but finite, temperature. Thus, we assume that the \emph{real} initial state of the system is the thermal state $\rho_0 = (e^{-\beta H}/Z)^{\otimes N}$ ($Z$ is the appropriate partition function) that reads explicitly:
\begin{equation}
    \rho_0 \equiv \left(
    \frac{1}{1+e^{-\beta\Delta E}}
    \begin{pmatrix}
    \displaystyle{e^{-\beta\Delta E}} & 0 \\
    0 & 1 \\
    \end{pmatrix}\right)^{\otimes N}
\end{equation}
where $\beta$ is the effective inverse temperature of the initial (prepared) state and $\Delta E$ is the energy difference between the single-qubit states $\ket{0}$ and $\ket{1}$. 

It is worth noting that the effective inverse temperature $\beta$ is not the \emph{actual} inverse temperature of the environment in which our quantum computer is located (albeit it will depend on it), but is a parameter that takes into account on average all the sources of disturbance that prevent our system to be in a perfectly pure state. For this reason, we will refer to it as an \emph{effective} temperature. Our choice to take a global constant value for the effective inverse temperature $\beta$, instead of setting different inverse temperatures $\{\beta_{1},\ldots,\beta_{N}\}$ for each qubit, stems from considering the average error on the initialization of the target state $\sigma_0$ on all the $N$ considered qubits for sake of clarity. Thus, without loss of generality, we can consider an average effective temperature that, in turn, makes our model easier to interpret. Moreover, let us note that with this notation, in the limit of zero temperature ($\beta\rightarrow\infty$), the state $\sigma_0$ is recovered, while in the opposite limit of infinite temperature ($\beta\rightarrow 0$) one gets the maximally mixed state $\mathbb{I}_{2^N}/2^N$, where $\mathbb{I}_d$ is the identity matrix of dimension $d$. We also recall that, given two density matrices $\rho$ and $\sigma$, representing the states of a quantum system, the fidelity between them is defined as $\mathcal{F}(\rho,\sigma) = \left( {\rm Tr}\left[ \sqrt{\sqrt{\rho}\,\sigma\sqrt{\rho}} \right]\right)^2$ \cite{jozsa1994fidelity}.

After setting our notation and initial assumptions, we formally show how the fundamental limit imposed by Nernst’s principle to the quantum state initialization affects the scaling of quantum computers. Let us take a perfect (in the sense of noiseless) unitary transformation $U$ operating on an ensemble of $N$ qubits, such that $\rho_1 \equiv U\rho_{0}U^\dagger$ and $\sigma_1 \equiv U\sigma_{0}U^\dagger$ are the resulting density operators after the application of the transformation. Then, we can find the analytical expression for the fidelity $\mathcal{F}(\rho_1,\sigma_1)$ as a function of the parameters $N$ and $\beta$. Since the fidelity is invariant under \emph{any} unitary transformations \cite{miszczak2008sub,nielsen2002simple} and $\sigma_0$ is a pointer state, we can prove that
\begin{equation}
    \mathcal{F}(\rho_1,\sigma_1)=\mathcal{F}(\rho_0,\sigma_0)=\text{Tr}\left[ \rho_0\sigma_0 \right].
    \label{eq:fidelity}
\end{equation}
By substituting the explicit form of $\rho_0$ and $\sigma_0$ in Eq.(\ref{eq:fidelity}), the scaling of $\mathcal{F}(\rho_1,\sigma_1)$ as a function of the parameters $N$ and $\beta$ is
\begin{equation}\label{eq:scaling}
    \mathcal{F}(\rho_1,\sigma_1) = \left( 1+e^{-\beta\Delta E} \right)^{-N}
\end{equation}
that is valid independently on which unitary transformation $U$ is applied. Eq.\,(\ref{eq:scaling}) shows that, even having at disposal any perfect unitary transformations $U$, a value slightly bigger than zero for the initial inverse temperature $\beta$ of the real state $\rho_0$ can end up hindering the scaling (i.e., $N\rightarrow\infty$) of the considered quantum circuit or algorithm. The reason behind this result being so general lies again in the thermodynamic considerations behind the Nernst’s unattainability principle, and thus in the divergent cost of attaining a perfect pure state (i.e., with $\beta\rightarrow\infty$). In fact, it now becomes clear that the issue of scaling quantum computers regards two competing limits:
\begin{align}
&\lim_{N\rightarrow\infty}\lim_{\beta\rightarrow\infty}\mathcal{F}(\rho_0, \sigma_0)=1\label{eq:limits_1} \\
&\lim_{\beta\rightarrow\infty}\lim_{N\rightarrow\infty}\mathcal{F}(\rho_0,\sigma_0)=0 \,.\label{eq:limits_2}
\end{align}
The non-commutativity of limits is ubiquitous in statistical physics and thermodynamics \cite{Geisel1995,Beekman2019,GherardiniGiachettiPRE,Buffoni2022}, where limits taken to infinity are often relevant and often related to mechanism of spontaneous symmetry breaking. In our case, Eq.\,(\ref{eq:limits_1}) states simply that if one is able to initialize a qubit in a pure quantum state, then in principle a \emph{perfect}, arbitrarily large quantum register can be realized. While, Eq.\,(\ref{eq:limits_2}) reflects the evidence that, for finite temperature, increasing the size of the quantum device necessarily entails a decrease in the attainable initial state fidelity $\mathcal{F}(\rho_0,\sigma_0)$ that will eventually disrupt the computation. The non-commuting nature of the two limits (\ref{eq:limits_1}) and (\ref{eq:limits_2}) is the second result of this work. In addition, the results of Eq.\,(\ref{eq:fidelity}) and Eq.\,(\ref{eq:scaling}) remain valid even if the real initial state $\rho_0$ contains residual quantum coherence (in the form of off-diagonal terms) originated by non-ideal state initialization routines. Refer to the Supplemental Material (SM) for the proof of this further result. Accordingly, we need to prepare pure states with increasing fidelity by properly taking into account also the needed resources, at least at the energetic level. In doing this, the initialization of quantum registers would need to be improved at a faster rate than the one at which the size of quantum computers increases. 

Let us now analyze the more general case of preparing a generic pure quantum state $|\psi\rangle$ from initializing the system in the real state $\rho_0$. Here, two kind of errors have to be considered: one on the initialization of $\sigma_0$ that is quantified by the fidelity $\mathcal{F}_{I} \equiv \mathcal{F}(\rho_0,\sigma_0)$, and the other on the subsequent preparation of $|\psi\rangle$. For the latter, the source of error stems from the fact the unitary operator $U$, needed for the preparation of $|\psi\rangle$, will be subjected to environmental noise in the form of a non-unitary quantum map $\Phi$. This second error is quantified by the fidelity 
\begin{equation}
\mathcal{F}_{P} \equiv 
\mathcal{F}(s_1,|\psi\rangle\!\langle\psi|)
= \langle 0|U^{\dagger}\Phi(\sigma_0)U|0\rangle
\end{equation}
where $s_1 \equiv \Phi(\sigma_0)$ and $|0\rangle \equiv \ket{00...0}$ for convenience. Hence, our aim is to have some information on the scaling of the fidelity $\mathcal{F}(\rho_1,|\psi\rangle\!\langle\psi|) = \langle 0|U^{\dagger}\Phi(\rho_0)U|0\rangle$ as a function of $N$, with $\rho_1 \equiv \Phi(\rho_0)$. The quantity $\mathcal{F}(\rho_1,|\psi\rangle\!\langle\psi|)$ is thus the fidelity of the composite process ``initialization + preparation''. As demonstrated in the SM, $\mathcal{F}(\rho_1,|\psi\rangle\!\langle\psi|)$ is bounded from below and above by functions depending only on $\mathcal{F}_{I}$ and $\mathcal{F}_{P}$:
\begin{equation}\label{eq:bounds_preparation_fid}
\mathcal{F}_{P}\mathcal{F}_{I} \leq \mathcal{F}(\rho_1,|\psi\rangle\!\langle\psi|) \leq \min\lbrace\mathcal{F}_{P},\mathcal{F}_{I}\rbrace \,.
\end{equation}
As a result, irrespective of the gate fidelity $\mathcal{F}_{P}$ whose computation requires some knowledge of $\Phi$, both the lower and upper bounds of $\mathcal{F}(\rho_1,|\psi\rangle\!\langle\psi|)$ scale at least with $\mathcal{F}_{I} = (1+e^{-\beta\Delta E})^{-N}$, namely exponentially in $N$. This is the third main result of this paper, which should further clarify the impact of the Nernst’s unattainability principle for the preparation of a pure quantum state in a real setting.

To reinforce the result of Eq.\,(\ref{eq:bounds_preparation_fid}), we also explicitly show that the initialization fidelity $(1+e^{-\beta\Delta E})^{-N}$ remains an upper bound to the attainable fidelity $\mathcal{F}(\rho_1,\sigma_1)$ of the final computation in case the non-unitary map $\Phi$ is a depolarizing quantum channel.
Depolarizing channels are commonly used to model noisy quantum circuits \cite{knill2005quantum, georgopoulos2021modelling}. Specifically, it can be proved (see the SM for the proof) that $\mathcal{F}(\rho_1,\sigma_1) \leq (1+e^{-\beta\Delta E})^{-N}$, where $\rho_1 = \Phi(U\rho_{0}U^{\dagger})$ and $\sigma_1 = U\sigma_{0}U^{\dagger}$ as above.

\begin{figure}
    \centering
    \includegraphics[width=0.95\linewidth]{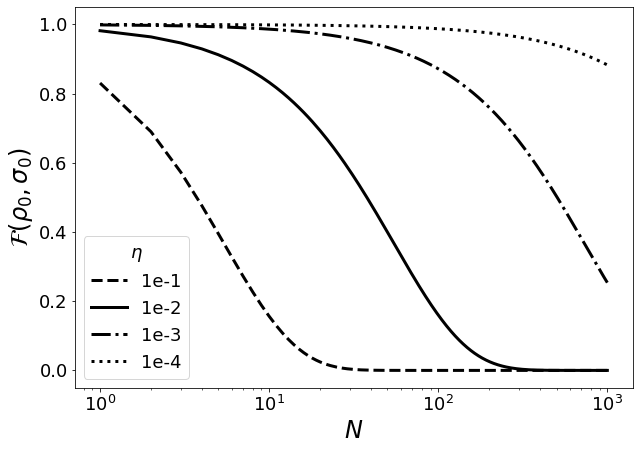}
    \caption{Scaling of the fidelity $\mathcal{F}(\rho_0,\sigma_0)$ as a function of the number $N$ of qubits (in log-scale), as predicted by Eq.\,(\ref{eq:scaling}), for different single qubit error rates $\eta=1-(1+e^{-\beta\Delta E})^{-1}$.}
    \label{fig:scaling}
\end{figure}

To better understand our results, we provide a quantitative gauge of the attainable precision (in terms of the fidelity function) of quantum computing, given a nonzero temperature of the initial qubit states. In this regard, in Fig.\,\ref{fig:scaling} one can observe a plot of the fidelity $\mathcal{F}(\rho_0,\sigma_0)$ with respect of the size $N$ of a qubit register for some values of the single qubit error rates $\eta$. The latter is related to the effective temperature $\beta$ by means of the relation $\eta \equiv 1-(1+e^{-\beta\Delta E})^{-1}$. In Fig.\,\ref{fig:scaling} it is apparent a sharp decay of the fidelity $\mathcal{F}(\rho_0,\sigma_0)$ while increasing the number $N$ of qubits: even by starting with quite accurate single qubit initialization, the fidelity will eventually start degrading.   

Once realized that perfect initialization may be challenging due to strict thermodynamic constraints imposed by the third law of thermodynamics, one shall necessarily perform quantum state initialization with an error \emph{good enough} to ensure that the fidelity $\mathcal{F}(\rho_0,\sigma_0)$ -- as provided by Eqs.\,(\ref{eq:fidelity}) and (\ref{eq:scaling}) -- is equal to the target value required for the operation. To put this into perspective, in order to have a target fidelity of $90\%$ for a quantum computer of $1000$ qubits, the error on the single qubit initialization has to be well below $10^{-4}$ that, to our knowledge, is the best recorded value \cite{Myerson2008,Burrell2010}.

\paragraph{Real data.--}

In this section we aim to understand in quantitative terms how the fidelity of current quantum devices scales as a function of the system's size and in relation to the quantum state initialization. For this purpose, test runs are performed using a superconducting quantum computer provided by IBM \cite{ibmq}. Specifically, our tests are run on the ibm-lagos quantum computer that, with 7 qubits and a quantum volume \cite{cross2019validating} of 32, was the larger device at our disposal. 

\begin{figure}[t]
    \centering
    \includegraphics[width=0.95\linewidth]{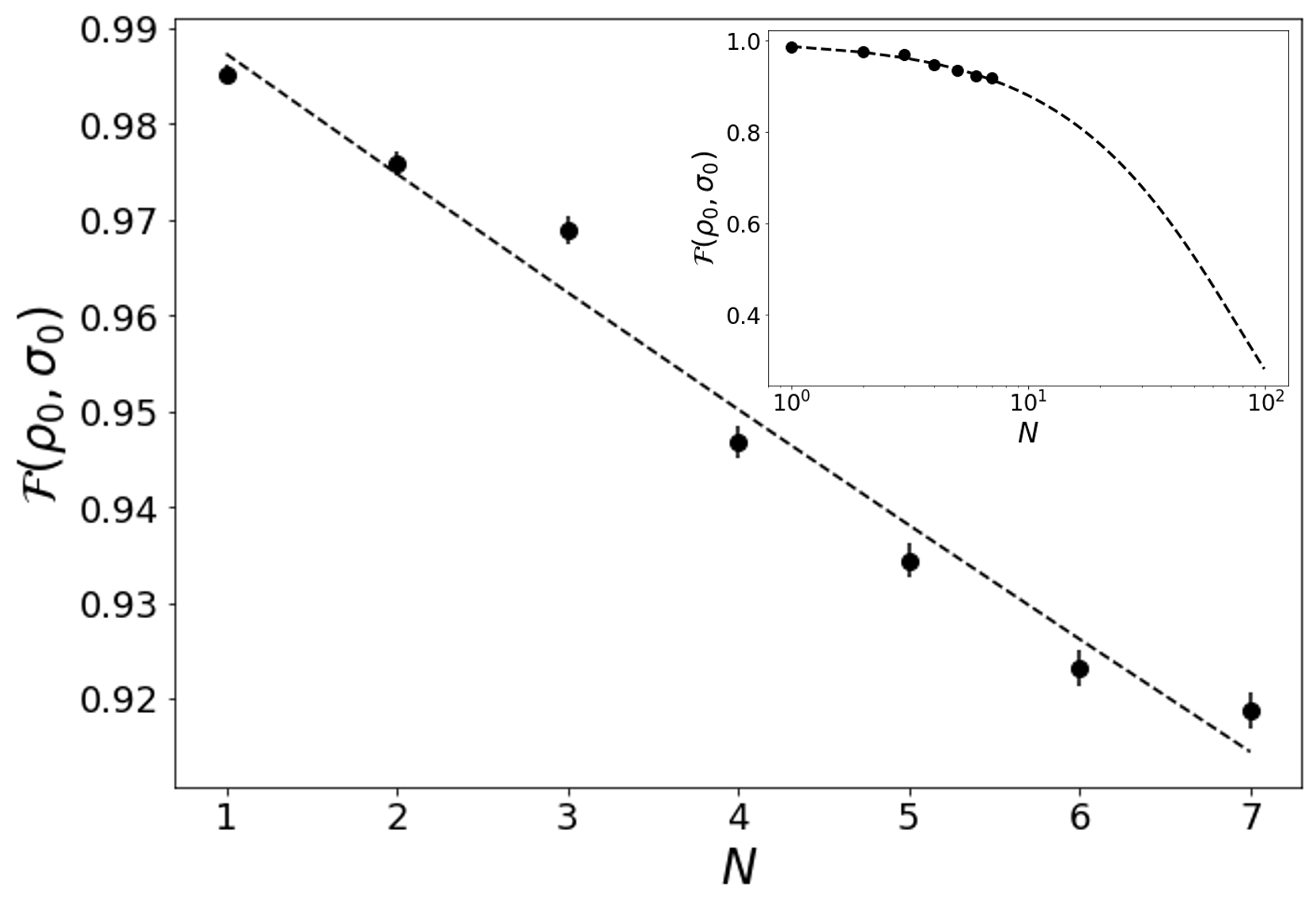}
    \caption{Measured fidelity values of the state $\sigma_0$ on the ibm-lagos quantum computer as a function of the number $N$ of reset qubits. The dashed line denotes the theoretical fit on the measured data using Eq.(\ref{eq:scaling}). In the inset, we show the same plot but with the theoretical fit extended to $100$ qubits; the log-scale on the x-axis is used for visualization purposes. Error bars over the measured values are computed by assuming that error fluctuations follow a Gaussian distribution.}
    \label{fig:exp-scaling}
\end{figure}

The first realized scaling protocol consists in locally measuring the initial register state $\ket{00...0}$ immediately after its preparation. For each value of $N$, the protocol is repeated $5 000$ times to collect statistics. In this implementation, the fidelity $\mathcal{F}(\rho_0,\sigma_0)$ is equal to the frequency by which $\ket{00...0}$ is measured. By performing the protocol for different number $N$ of qubits, we obtain the results reported in Fig.\,\ref{fig:exp-scaling}.
From the figure one can observe that, while the single-qubit initialization fidelity is almost $99\%$, as the number of qubits increases this value drops significantly to around $92\%$. For a quantitative evaluation, we assume the fidelity to be scaling as Eq.(\ref{eq:scaling}), and we fit the value of $\beta\Delta E$ over the measured data, getting a value of $\beta\Delta E = 4.35\pm0.03$ with a coefficient of determination $R^2=0.976$. The resulting curve, whose analytical expression is provided by Eq.\,(\ref{eq:scaling}), is plotted as the dashed line in Fig.\,\ref{fig:exp-scaling}. Since IBM provides us with the values of $\Delta E$ \cite{ibmq} for each qubit of the employed processor (all around $5$ GHz), we can thus compute the effective temperature $\beta$ that for the realized test is placed at $56.80\pm 1.21$ mK. This effective temperature, as expected, is a bit higher than the physical temperature of the fridge ($\sim 15$ mK), since it takes into account also the effect of sources of noise such as measurement-induced or single-gate errors. Note that our choice to take a global constant for the effective temperature gives a small discrepancy between the measured values and the theoretical scaling curve, albeit in the real device every qubit should have its own temperature. The trend of the fidelity scaling provided by Eq.\,(\ref{eq:scaling}), versus the size $N$, is shown in the inset of Fig.\,\ref{fig:exp-scaling} where the predicted fidelity is evaluated for a circuit composed by a larger number of qubits. We remark here that, since the number of qubits at our disposal was just up to $N=7$, this scaling is an extrapolation from our theory and the fit we provide still does not constitute a definitive proof that the scaling we propose is indeed the correct one (it serves as a visual guide of what our bounds predict). However, the fidelity scaling predicted by our theory can be also applied on already published and available data, e.g.~for the preparation of GHZ states on trapped ion platforms~\cite{MonzPRL2011,PogorelovPRXQuantum2021} and the sampling from random circuits with fixed depth~\cite{AruteNature2019} on a superconducting chip. As discussed in detail in the SM, we can use our model to both extrapolate useful information and fit the fidelity values in \cite{PogorelovPRXQuantum2021,AruteNature2019} as a function of the number of qubits $N\in[12,53]$. This provides a further validation of our model beyond the 7-qubits regime. Hence, from the results in Fig.\,\ref{fig:exp-scaling}, it becomes evident the so fundamental role played by quantum state initialization and the effective temperature for the realization of a large-scale quantum computer (or more generally a quantum device), before its fidelity dramatically decreases.

\paragraph{Reset protocols.--}

We now focus on understanding how these fidelity values can be improved. For such a purpose, active reset methods have been devised, which fall into two categories: conditional \cite{Walter2017,Riste2012b,Govia2015} and unconditional \cite{Geerlings2013,Egger2018,Magnard2018, goldberg2021breaking} resets. We here employ a mixture of conditional resets methods and thermalization inspired by other works performed by IBM \cite{ibmreset} where we take a register of qubits, initially prepared in the superposition state $\ket{+}^{\otimes N} \equiv H^{\otimes N}\ket{00...0}$
(with $H$ being the Hadamard gate)\,\footnote{We could also use the results of any previous quantum computation quantum computation leaving the processor in a state $\ket{\psi}$ and setting the variable \texttt{init\_qubits=false}} and reset to $\ket{00...0}$ by means of $K$ consecutive conditional resets. In this conditional reset protocol, each qubit of the register is measured and then a ${\rm NOT}$-gate is applied conditionally on the measurement outcome. Ideally, the register is reset to $\ket{00...0}$ with zero error, but practically its state is set to the density operator $\rho_K$. 
\begin{figure}[t!]
    \centering
    \includegraphics[width=0.95\linewidth]{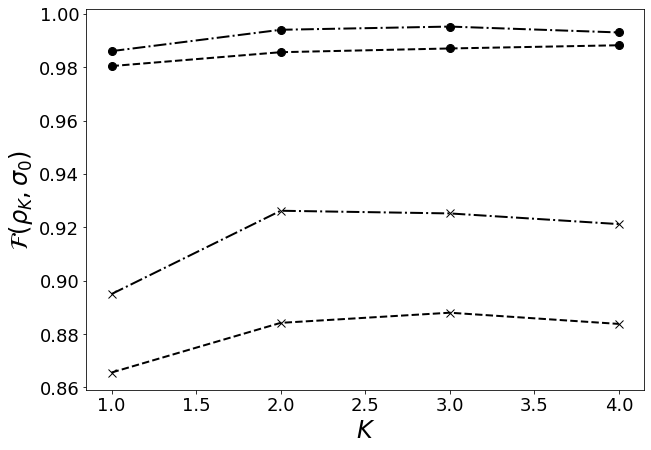}
    \caption{Measured fidelity between the quantum computational state $\sigma_0$ and the density operator $\rho_K$, solution of the conditional reset protocol, as a function of $K$. The circle markers represent conditional resets on a single-qubit register, while the cross markers represent the conditional resets on a $7$-qubits register. Dashed lines refer to consecutive resets without delay and dash-dotted lines to resets with a delay of $500 \mu s$ between them. The error bars are smaller than the size of the markers.}
    \label{fig:multiple-resets}
\end{figure} 

In Fig.\,\ref{fig:multiple-resets}, the results of the conditional reset runs, carried out on the $7$-qubits IBM quantum computer ibm-lagos , are plotted for a varying number of resets $K$. As one can note, by increasing the number of resets (i.e., by employing more energy to carry out the reset protocol), the state reset fidelity increases up to a certain plateau, whose value depends on {\it (i)} the measurement readout error, {\it (ii)} the gate noise affecting the ${\rm NOT}$ operation, as well as {\it (iii)} the thermalization of the qubit due to the external environment. We also observe that we can further increase the fidelity of our reset protocol by inserting a delay of $500\mu s$ between two consecutive reset (dash-dotted lines in Fig.\,\ref{fig:multiple-resets}) during which the qubits thermalize with the environment. This delay value of $500\mu s$ is the maximum that we could implement on our machine and we observed that in general, for different values, increasing the delay leads to a better reset fidelity. A similar behaviour was also observed in \cite{blogpost} for a range of different processors. Moreover, one can see that the difference between the results provided by the reset protocol applied both to a single-qubit register and to the $7$-qubits one lies in the plateau's value (that decreases as the size of the register increases) and not in the number $K$ of resets needed to reach the maximum allowed fidelity. Our findings are the hint that achieving greater fidelity values amounts to expending a larger amount of thermodynamic resources in the state initialization protocol, as we quantified in Fig.\,\ref{fig:multiple-resets}.

\paragraph{Conclusions.--}

To conclude, we investigated how the fidelity of initializing a quantum register, as well as preparing a generic pure quantum state, is constrained by the statement of the third law of thermodynamics. The expected scaling follows the one expressed by Eq.\,(\ref{eq:scaling}) for every finite value of the effective inverse temperature $\beta$ of the initial qubits register. We also observe a scaling compatible to the one in Eq.\,(\ref{eq:scaling}) on a real quantum computer, albeit limited to 7-qubits. The solution to the challenge posed by this constraint is to use better protocols and using more resources in order to reach the target fidelity values needed given the desired size of the quantum register.

In this regard, it would be of great interest to investigate implementations of conditional and unconditional resets, as well as ideas which avoid resets entirely \cite{Werninghaus2021}. A detailed study of all the variants of qubit reset is timely and of great importance to the future of quantum computing. In future investigations, one could also explore if quantum computing can be redesigned to operate (even partially) on mixed quantum states \cite{siomau2011quantum}. In conclusion, we would like to stress the relevance that the thermodynamic study of quantum systems will have for the development of quantum devices and the successful realization of large-scale quantum computers \cite{cimini2020experimental,aifer2022quantum,auffeves2021quantum,Cafaro2022complexity}. As we showed in this work, considerations about energy dissipation, finite-temperature states and other thermodynamic quantities will be key aspects for the next developments in practical applications of quantum computing.

\begin{acknowledgments}
\paragraph{Acknowledgments.--} We gratefully thank Per Delsing, Thomas Monz, and Michele Campisi for pointing out useful references and fruitful discussions. We also acknowledge the access to advanced services provided by the IBM Quantum Researchers Program. The views expressed are those of the authors, and do not reﬂect the official policy or position of IBM or the IBM Quantum team. E.Z.C. and Y.O. thank the support from FCT -- Funda\c c\~ao para a Ci\^encia e a Tecnologia (Portugal), namely through project UIDB/50008/2020 and UIDB/04540/2020, as well as from projects TheBlinQC and QuantHEP supported by the EU H2020 QuantERA ERA-NET Cofund in Quantum Technologies and by FCT (QuantERA/0001/2017 and QuantERA/0001/2019, respectively), and from the EU H2020 Quantum Flagship project QMiCS (820505). S.G. acknowledges The Blanceflor Foundation for financial support through the project ``The theRmodynamics behInd thE meaSuremenT postulate of quantum mEchanics (TRIESTE)''.
\end{acknowledgments}

\bibliographystyle{unsrt}
\bibliography{bibliography}

\appendix*
\section*{Proof of Eq.(3)}

The fidelity between two generic density matrices $\rho$ and $\sigma$ is provided by the well-known Uhlmann fidelity that is defined as $\mathcal{F}(\rho,\sigma) = \left( {\rm Tr}\left[ \sqrt{\sqrt{\rho}\,\sigma\sqrt{\rho}} \right]\right)^2$. If at least one of the two states is pure, i.e., $\sigma = |\psi\rangle\!\langle\psi|$, the expression of $\mathcal{F}(\rho,\sigma)$ simplifies as $\mathcal{F}(\rho,\sigma) = \text{Tr}\left[\rho\sigma\right]$. Let us prove such a statement. The proof is based in the evidence that the fidelity $\mathcal{F}$ is invariant under unitary transformations \cite{miszczak2008sub}, i.e.,
\begin{equation}
    \mathcal{F}(U\rho_0\,U^\dagger,U\sigma_0\,U^\dagger) = \mathcal{F}(\rho_0,\sigma_0)
\end{equation}
where $U$ generic unitary operator. This entails that, by defining $\rho_1 \equiv U\rho_0\,U^\dagger$ and $\sigma_1 \equiv U\sigma_0\,U^\dagger$, the following equation holds: 
\begin{equation}\label{app_first_formula}
    \mathcal{F}(\rho_1,\sigma_1) = \mathcal{F}(\rho_0,\sigma_0) = \text{Tr}\left[ \rho_0\sigma_0 \right].
\end{equation}
Eq.\,(\ref{app_first_formula}) is exactly Eq.(3) in the main text.

\section*{Proof of Eq.\,(4) by including initial quantum coherence}

Let us now assume that the actual initial state of the considered quantum system is not just
\begin{equation}
    \rho_0 \equiv  
    \left(\frac{1}{1+e^{-\beta\Delta E}}
    \begin{pmatrix}
    \displaystyle{e^{-\beta\Delta E}} & 0 \\
    0 & 1 \\
    \end{pmatrix}\right)^{\otimes N},
\end{equation}
thermal state at inverse temperature $\beta$, also contains quantum coherence terms that one may consider as a defect of the reset protocol for quantum state initialization. Hence, let us consider that the actual initial state (after the initialization procedure) is
\begin{equation}
    \overline{\rho}_0 \equiv \left( 
    \frac{1}{1+e^{-\beta\Delta E}}
    \begin{pmatrix}
    \displaystyle{e^{-\beta\Delta E}} & \epsilon \\
    \epsilon^* & 1 \\
    \end{pmatrix}\right)^{\otimes N}.
\end{equation}
If, then, we make use of Eq.(3), one simply gets that 
\begin{equation}
\mathcal{F}(\overline{\rho}_1,\sigma_1) = \text{Tr}\left[ \overline{\rho}_0\sigma_0 \right].
\end{equation}
Afterwards, since $\sigma_0 \equiv |00...0\rangle\!\langle 0...00|$ is a projector on the computational basis, we still have that
\begin{equation}
    \mathcal{F}(\overline{\rho}_1,\sigma_1) = \left( 1+e^{-\beta\Delta E} \right)^{-N}.
\end{equation}
In conclusion, Eq.\,(4) remains valid even for initial states which are not purely thermal (or mixed) ones, i.e, in other terms, initial quantum coherence in the register qubit states does not affect the scaling of the fidelity $\mathcal{F}(\rho_1,\sigma_1)$.

\section*{Proof of Eq.\,(8)}

Let us consider the problem of preparing the generic pure state $|\psi\rangle$ by applying a generic unitary $U$ to $|00...0\rangle$. 

Here, we prove the upper and lower bounds for the fidelity $\mathcal{F}(\rho_1,|\psi\rangle\!\langle\psi|) \equiv \langle\psi|\rho_1|\psi\rangle$ concerning both the initialization of $|00...0\rangle$ and the subsequent preparation of $|\psi\rangle$, where $|\psi\rangle \equiv U|00...0\rangle$ and $\rho_1 \equiv \Phi(\rho_0)$ with $\Phi$ a non-unitary quantum map and 
\begin{equation*}
    \rho_0 \equiv  
    \left(\frac{1}{1+e^{-\beta\Delta E}}
    \begin{pmatrix}
    \displaystyle{e^{-\beta\Delta E}} & 0 \\
    0 & 1 \\
    \end{pmatrix}\right)^{\otimes N}
\end{equation*}
as above. This state-preparation routine (starting from a register with $N$-qubits), even if performed from the ideal state $\sigma_0 \equiv |00...0\rangle\!\langle 0...00|$ will generate in itself some errors due to the fact that quantum operations are not perfect and the system is in contact with the external environment. Thus, in general we are interested in knowing how this preparation error (depending on the circuit that implements the unitary $U$) combines with the initialization error (depending on the inverse temperature $\beta$), so as to obtain the fidelity of the prepared state with respect to the ideal case.

\begin{figure*}[t]
    \centering
    \includegraphics[width=0.95\linewidth]{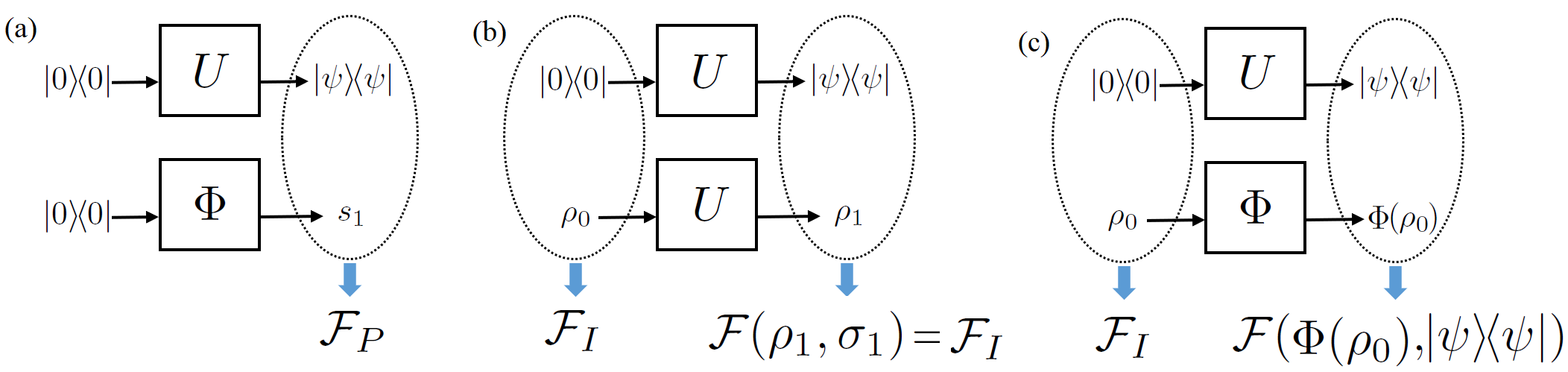}
    \caption{Pictorial representation of the physical settings that define the fidelities $\mathcal{F}_{P}$, $\mathcal{F}_{I}$ and $\mathcal{F}(\rho_1,|\psi\rangle\!\langle\psi|)$, respectively in panels (a)-(c).}
    \label{fig:schemes}
\end{figure*}

Let us first introduce some assumptions. In general, by supposing to start from the perfectly initialized state $\sigma_0$, the \emph{imperfect} preparation process can be modelled as a quantum map expressed in terms of the set of Kraus operators $\{K_i\}$. As a result, we define $s_1 \equiv \sum_i K_i\sigma_{0}K^{\dagger}_i$
such that the fidelity of the preparation of $|\psi\rangle$ by starting from the initial register state $|00...0\rangle$ is (see also panel (a) of Fig.\,\ref{fig:schemes})
\begin{equation}
    \mathcal{F}_{P} \equiv 
    \mathcal{F}(s_1,|\psi\rangle\!\langle\psi|)=\sum_{i}|\langle\psi|K_i|0\rangle|^{2},
\end{equation}
where from here on $|00...0\rangle$ is denoted for simplicity as $|0\rangle$ implying that we are always working with a register of $N$ qubits. Then, let us also recall the fidelity $\mathcal{F}_{I}$ in initialising the initial state $|0\rangle$ of the register. Specifically, 
\begin{eqnarray}
    \mathcal{F}_{I} \equiv \mathcal{F}(\sigma_0,\rho_0) =  \left( 1+e^{-\beta\Delta E} \right)^{-N}
\end{eqnarray}
that is the main topic of this work (panel (b), Fig.\,\ref{fig:schemes}). 

We now turn our attention on the composite process consisting on initialization and preparation of the generic pure quantum state $|\psi\rangle$. Let us thus consider $\rho_1 = \Phi(\rho_0) = \sum_{i}K_i\rho_{0}K^{\dagger}_i$, as illustrated in panel (c) of Fig.\,\ref{fig:schemes}. Then, the fidelity of the composite process can be computed as
\begin{equation}\label{eq:fidelity_composite_process}
\mathcal{F}(\rho_1,|\psi\rangle\!\langle\psi|) = \sum_{i}\bra{\psi} K_i\rho_{0}K^{\dagger}_i\ket{\psi}.
\end{equation}
If we insert the identity operator $\mathbb{I}=\sum_{k}|k\rangle\!\langle k|$ both on the left and the right of $\rho_0$ in Eq.\,(\ref{eq:fidelity_composite_process}), one then gets
\begin{equation}\label{eq:fidelity_composite_process_2}
 \mathcal{F}(\rho_1,|\psi\rangle\!\langle\psi|)= \sum_i\sum_{m,n}\bra{\psi} K_i\ket{m}\bra{m}\rho_0\ket{n}\bra{n}K^{\dagger}_i\ket{\psi}.
\end{equation}
In this regard, it is worth noting that $\{|k\rangle\}$ is the set containing all the elements of the computational basis that naturally decompose the $N$-qubits states of the quantum circuit/computer we are taking into account. Now, since $\rho_0$ is a diagonal operator, the only nonzero terms in the summations of Eq.\,(\ref{eq:fidelity_composite_process}) are the ones with $m=n$. This simplifies Eq.\,(\ref{eq:fidelity_composite_process_2}) as
\begin{equation}\label{eq:fidelity_composite_process_3}
 \mathcal{F}(\rho_1,|\psi\rangle\!\langle\psi|)= \sum_n\sum_i|\bra{\psi} K_i\ket{n}|^2\bra{n}\rho_0\ket{n}.
\end{equation}
Observe that the right-hand-side of Eq.\,(\ref{eq:fidelity_composite_process_3}) is given by the sum over positive elements in $n$; thus, if we take just the term $n=0$, it holds that
\begin{equation}\label{eq:fidelity_composite_process_4}
 \mathcal{F}(\rho_1,|\psi\rangle\!\langle\psi|)\geq \sum_i|\bra{\psi} K_i|0\rangle|^{2}\langle 0|\rho_0|0\rangle=\mathcal{F}_{P}\mathcal{F}_{I}\,.
\end{equation}

Previously, we have shown that $\mathcal{F}_{I}$ always holds as an upper bound to the fidelity $\mathcal{F}(U\rho_{0}\,U^{\dagger},U\sigma_{0}\,U^{\dagger})$ for the property of invariance of $\mathcal{F}$ under unitary transformations. Hence, in the general case whereby $\Phi$ is an arbitrary non-unitary channel, $\mathcal{F}(\rho_1,|\psi\rangle\!\langle\psi|) \leq \mathcal{F}_{I}$. However, in general, $\mathcal{F}(\rho_1,|\psi\rangle\!\langle\psi|)$ is also smaller or equal than $\mathcal{F}_{P}$. In fact, if $\rho_0 = \sigma_0$, then $\mathcal{F}_{I}=1$ (thus, no initialization error) and $\mathcal{F}(\rho_1,|\psi\rangle\!\langle\psi|)=\sum_{i}|\langle\psi|K_i|0\rangle|^{2}=\mathcal{F}_{P}$. Therefore, overall, the individual fidelities for the initialization of $|0\rangle$ and the preparation of $|\psi\rangle$ can hold as upper bounds of $\mathcal{F}(\rho_1,|\psi\rangle\!\langle\psi|)$, with the result that for a generic state preparation we have:
\begin{equation}
\mathcal{F}_{P}\mathcal{F}_{I} \leq\mathcal{F}(\rho_1,|\psi\rangle\!\langle\psi|)\leq \min\lbrace\mathcal{F}_{P},\mathcal{F}_{I}\rbrace \,.
\end{equation}

It thus remains to address the issue of characterizing the state preparation fidelity $\mathcal{F}_{P}$. By construction, the latter depends on the way the pure quantum state $|\psi\rangle$ is prepared (thus, it depends on the specific quantum circuit we have chosen for the state preparation) and on the average error made in realizing such quantum circuit. In this regard, note that $\mathcal{F}_{P}$ is identically equal to the gate fidelity \cite{nielsen2002simple} 
\begin{equation}\label{eq:gate_fidelity}
    \langle 0|U^{\dagger}\Phi(\sigma_0)U|0\rangle \,.
\end{equation}
A universal form of the scaling of Eq.\,(\ref{eq:gate_fidelity}) as a function of $N$, number of qubits, is still unknown and it is outside of the scope of our work. Nevertheless, what we are able to state about the preparation of a generic pure quantum state, as novel result, is the way the fidelity $\mathcal{F}(\rho_1,|\psi\rangle\!\langle\psi|)$ of the composite process ``initialization + preparation'' is bounded from above and below by a quantity that scales at least as $\mathcal{F}_{I}$. Accordingly, irrespective of the gate fidelity $\mathcal{F}_{P}$ whose computation requires some knowledge of $\Phi$, both the lower and upper bounds scale at least exponentially with $N$.

\section*{Depolarizing quantum channel as noise modelling}

We here prove that $\left( 1+e^{-\beta\Delta E} \right)^{-N}$ is still the \emph{upper bound} to the attainable fidelity also in the case the quantum gate is also followed by a depolarizing channel $\mathcal{E}$ acting on $\rho$ as $\mathcal{E}(\rho) \equiv (1-\lambda)\rho + \lambda{\rm Tr}[\rho]\mathbb{I}/2^{N}$, with $\rho$ a generic density operator and $\mathbb{I}$ the $2^N$ dimensional identity matrix. The depolarizing channel is a model for quantum errors commonly affecting quantum systems in general and quantum gates in particular \cite{KuengPRL2016,knill2005quantum}, which makes non-unitary the applied quantum operation. 

Let us thus apply the depolarizing channel $\mathcal{E}$ to $U\rho_{0}\,U^{\dagger}$; one gets:
\begin{equation}
\rho_1 = \mathcal{E}(U\rho_{0}\,U^\dagger) = (1-\lambda)U\rho_{0}\,U^\dagger + \frac{\lambda}{2^N}\text{Tr}[U\rho_{0}\,U^\dagger]\,\mathbb{I},
\end{equation}
where $\lambda$, with $0\leq\lambda\leq 4^{N}/(4^{N} - 1)$, is the parameter that quantifies how much the channel is non-unitary. Our target state $\sigma_1$, on the other hand, would be $\sigma_1=U\sigma_0 U^\dagger$. Then, we want to check whether
\begin{equation}
    \mathcal{F}(\rho_0,\sigma_0) = \text{Tr}[\rho_0\sigma_0] \geq \text{Tr}[\rho_1\sigma_1] = \mathcal{F}(\rho_1,\sigma_1).
\end{equation}
For this purpose, $\text{Tr}[\rho_1\sigma_1]$ can be explicitly expressed as $\text{Tr}[\rho_1\sigma_1] = (1-\lambda)\text{Tr}[\rho_0\sigma_0]+\frac{\lambda}{2^N}$ by using the cyclic property of the trace. In this way, one obtains the following inequality:
\begin{equation}\label{app_inequality}
    \text{Tr}[\rho_0\sigma_0] \geq (1-\lambda)\text{Tr}[\rho_0\sigma_0]+\frac{\lambda}{2^N}.
\end{equation}
With some simple manipulations, the inequality (\ref{app_inequality}) simplifies as
\begin{equation}
    \text{Tr}[\rho_0\sigma_0] \geq 2^{-N}.
\end{equation}
Finally, by recalling the explicit expression of $\text{Tr}[\rho_0\sigma_0]$, we have that
\begin{equation}
    \left( 1+e^{-\beta\Delta E} \right)^{-N} \geq 2^{-N}
\end{equation}
that is always true $\forall \beta \geq 0$. We have thus proved, for this specific noise model, the scaling of a generic state preparation.

\section*{Experiments on other available data}

\begin{figure}[ht]
    \centering
    \includegraphics[width=\linewidth]{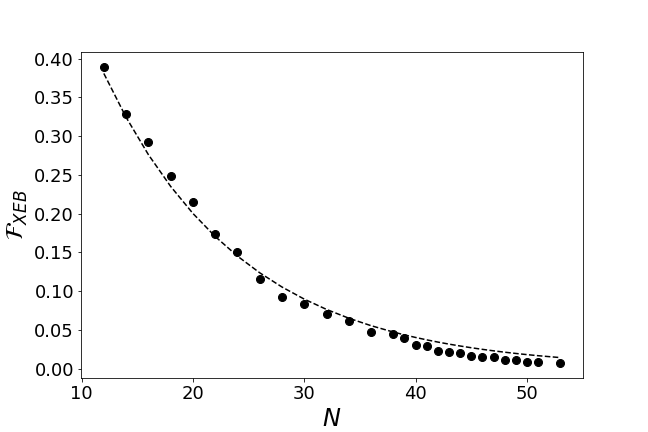}
    \caption{
    Data of the cross-entropy fidelity $\mathcal{F}_{XEB}$ as measured in the experiments of \cite{AruteNature2019} (Fig.4a) for a circuit with fixed depth. The dashed line represents the best fit of our theoretical prediction on the data, with a value of $\beta\Delta E = 2.48\pm 0.01$ and a coefficient of determination for the fit of around $R^2=0.99$.
    }
    \label{fig:google}
\end{figure}

In order to extrapolate our theoretical findings beyond the 7-qubit regime as in our runs on the IBM platform, we are now going to use our theoretical bound to explain the scaling of data already existing in the current literature. For example, in the celebrated paper about quantum supremacy by the research quantum laboratory of Google \cite{AruteNature2019}, the metric used to benchmark the implemented (random) circuits is a fidelity, denoted as \emph{cross-entropy fidelity} by the authors. Thus, we can use the available data provided in \cite{AruteNature2019} to have a further comparison with our model. In fact, since the depth of the circuits is fixed in the experiments in \cite{AruteNature2019}, we can model the cumulative effects of the noisy initialization of the circuit and the noise sources affecting the realization of the circuit itself just as a noisy initialization (with a higher temperature) of a noiseless circuit. Using this assumption, we can fit our theoretical scaling on their measured cross-entropy fidelity from a number of qubits ranging from $N=12$ to $N=53$. As one can observe in Fig.\ref{fig:google}, the scaling provided by our theoretical model fits the observed fidelity values of \cite{AruteNature2019} with very high accuracy. This extrapolation of our fidelity scaling to a larger number of qubits gives us more confidence on the validity of the temperature model we are here providing. Moreover, this also allows to model the effect of noise on fixed-depth circuits just by assuming an initial finite temperature to the initial state preparation.

A similar scaling has been also observed for the preparation of GHZ states, both theoretically \cite{PogorelovPRXQuantum2021} and experimentally \cite{MonzPRL2011} (in particular, we refer to Fig.\,17b) on a 24-qubits trapped ion platform. In order to get a GHZ state from the initial register state $\ket{00...0}$, the implementation of a M{\o}lmer-S{\o}renson gate is required. In such a case, the dominant effect explaining the fidelity decay is likely due to the number of operations needed to prepare the GHZ state, and thus to their error scaling that goes as $\sim N^2$. 
Following our results, we can thus postulate that the exponential decay of the GHZ state preparation fidelity as a function of $N$ has a finite-temperature contribution given by $(1+e^{-\beta\Delta E})^{-N}$, which can serve as an upper bound for the attainable fidelity. We estimate that for the values of $\eta\sim 5\cdot 10^{-3}$ reported in \cite{MonzPRL2011}, the fidelity to initialize the state $\ket{00...0}$ for the 24-qubits will be $\sim 90\%$, while the actual measured fidelity \emph{after} the circuit required to create the GHZ state is $\sim 50\%$. As final remark, note that for the preparation of GHZ states, contrary to Google's experiments, we cannot directly fit the scaling of the fidelity with our model since the depth of the circuit required to implement the GHZ state is not fixed but scales with $N$.

\end{document}